\documentclass{moriond}

\def\Journal#1#2#3#4{{#1} {\bf #2}, #3 (#4)}

\def\JCAP{{\em Journal of Cosmology and Astroparticle Physics}}

\def\be{\begin{equation}}
\def\ee{\end{equation}}
\def\bea{\begin{eqnarray}}
\def\eea{\end{eqnarray}}

\begin{document}
\vspace*{4cm}
\title{BOLOMETRIC INTERFEROMETRY AND SPECTRAL IMAGING: A QUBIC OVERVIEW}

\author{T.~LACLAVERE on behalf of QUBIC collaboration}

\address{Laboratoire Astroparticule et Cosmologie (APC), Université Paris-Cité, Paris, France}

\maketitle
\abstracts{The Q \& U Bolometric Interferometer for Cosmology (QUBIC) is a new kind of cosmological instrument that uses interferometry to observe the sky. The unique synthesized beam of QUBIC has a significant frequency dependence that we use to increase spectral resolution within the QUBIC bandwidth, allowing us to mitigate foreground contamination in an improved manner.}

\section{QUBIC Overview}
QUBIC is an instrument located in Argentina, designed to detect B-modes of polarization in the Cosmic Microwave Background (CMB). The particularity of this instrument is to use interferometry combined with high-sensitivity bolometers.
The core of QUBIC is the back-to-back horn array, producing multiple beams optically combined by two convex mirrors to create an interference pattern on the focal planes. This array comprises 64 horns for the Technical Demonstrator (TD) and 400 for the Full Instrument~\cite{QUBIC VIII}.

\begin{figure}[h]
\centerline{
\includegraphics[width=0.8\linewidth]{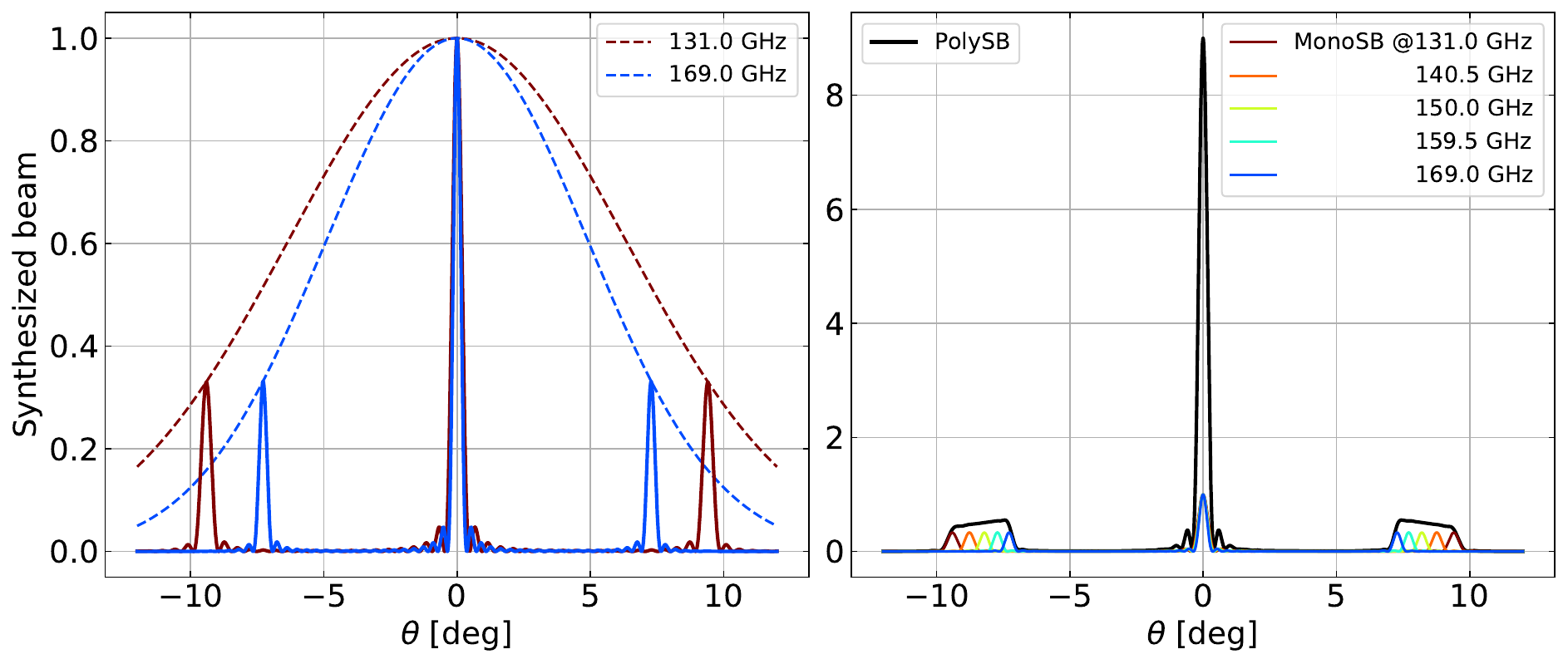}
}
\caption[]{1D cut of the QUBIC synthesized beam~\cite{QUBIC II} at two monochromatic frequencies (left) and corresponding instrument response for a broadband source (right).}
\label{fig: schema synthesized beam}
\end{figure}

\section{Spectral-Imaging}
The particularity of QUBIC, the horn array, determines the shape of the QUBIC Point-Spread-Function (PSF), which we call the synthesized beam. It is a superposition of interferometry fringe patterns, with fringe distance inversely proportional to frequency, exactly as in Young's experiment. The obtained beam, the sum of all fringes, is also frequency-dependent: the distance between the secondary and central peaks is inversely proportional to frequency. Figure~\ref{fig: schema synthesized beam} (left) shows a cut through this beam to highlight this effect. \\
The particularity associated with this synthesized beam is the possibility to perform spectral imaging. Observing a polychromatic source results in ``secondary plateaux'' instead of peaks, as in Figure~\ref{fig: schema synthesized beam} (right). The idea of spectral imaging is to bin these plateaux in sub-areas, each being related to a sub-frequency band smaller than the instrumental physical bandwidth, increasing our spectral resolution. This feature is particularly useful for foreground mitigation, especially in the case of complex foregrounds such as frequency decorrelated dust~\cite{Decorrelation}. \\
An important advantage of spectral imaging is that it is achieved at the data analysis stage, after data acquisition, enabling flexible sub-band division based on analytical needs.

\section{Results}
\begin{figure}[t]
\begin{minipage}{0.65\linewidth}
\centerline{\includegraphics[width=1\linewidth]{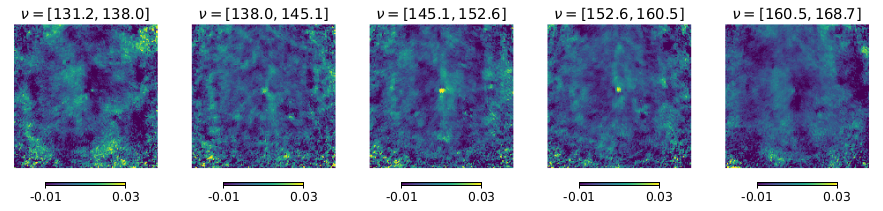}}
\end{minipage}
\hfill
\begin{minipage}{0.30\linewidth}
\centerline{\includegraphics[width=1\linewidth]{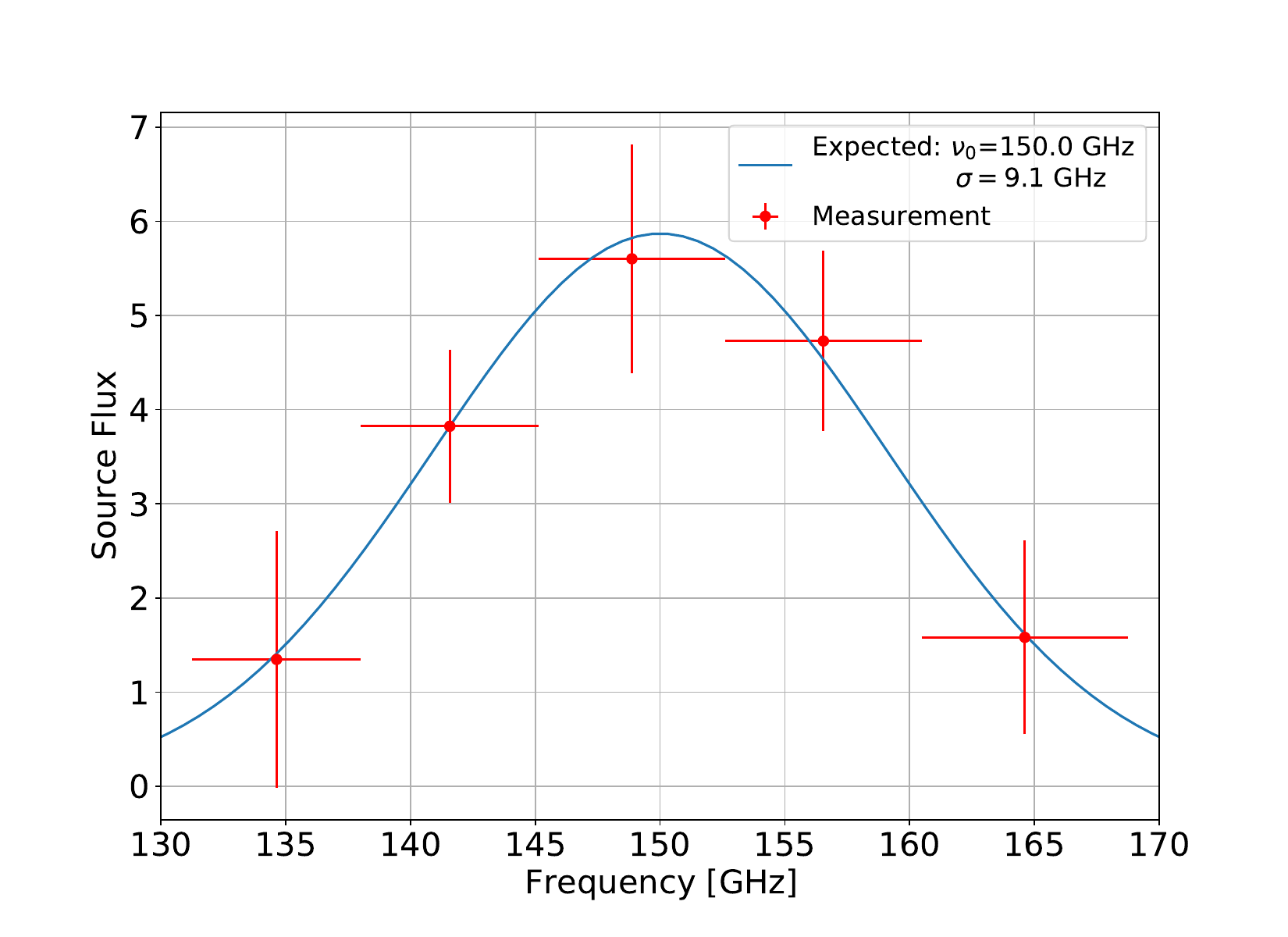}}
\end{minipage}
\hfill
\caption[]{(left) Calibration data with the source at 150 GHz projected on the sky when splitting the physical band into 5 sub-bands using spectral-imaging. (right) Measurement of the flux of the source in reconstructed sub-bands. The measurement in red is compared to the expected source spectrum convolved by the spectral resolution of QUBIC in blue}
\label{fig: results synthesized beam}
\end{figure}

In this section, we present an example of spectral imaging performed in the laboratory. We observed with the QUBIC instrument an artificial monochromatic source, emitting at 150 GHz (with 144 Hz bandwidth)~\cite{QUBIC III}. An imager observing over 130-170 GHz would dilute this signal by averaging within the band. With QUBIC, we can use spectral imaging to split this range into smaller sub-bands (Figure \ref{fig: results synthesized beam}). Thus, the monochromatic source can be observed in several frequency maps, we can then reconstruct its spectrum~\cite{QUBIC Moon}. By doing this, we were able to measure the emission of the source, limited by the QUBIC spectral resolution, as shown in Figure \ref{fig: results synthesized beam}. 
We foresee applications beyond component separation methods: applying this method to the 230 GHz Carbon Monoxide line. By mapping separately the data associated with this precise frequency using spectral imaging, we can reduce contamination in the rest of the band, enhancing foreground characterization. A study is underway to apply spectral imaging to mitigate atmospheric contamination in the TOD.

\end{document}